\documentclass[preprint,12pt, authoryear]{elsarticle}
\usepackage[utf8x]{inputenc}
\usepackage[T1]{fontenc}
\usepackage{mathptmx} 
\usepackage{amssymb} 
\usepackage{calc} 
\usepackage{enumitem} 
\usepackage[authoryear]{natbib}
\usepackage[a4paper, lmargin=0.1666\paperwidth, rmargin=0.1666\paperwidth, tmargin=0.1111\paperheight, bmargin=0.1111\paperheight]{geometry} 
\usepackage{color}

\usepackage[all]{nowidow} 
\usepackage[protrusion=true,expansion=true]{microtype} 
\usepackage{hyperref} 
\usepackage{graphicx} 
\usepackage{booktabs} 
\usepackage[para,online,flushleft]{threeparttable}
\usepackage{pdflscape}
\usepackage{tabularx}

\journal{Annals of the American Association of Geographers}

\begin{document}

\begin{frontmatter}



\title{Uncovering the Associations between Human Big Five Personality Traits and Built Environment Characteristics from Street View Imagery}

\author[doge,doa]{Koichi Ito}
\author[doge]{Yuhao Kang\corref{cor1}}
\author[dop]{Samuel D Gosling}
\author[doge]{Xihan Yao}
\author[atof]{Jeff Potter}
\author[doa,dre]{Filip Biljecki}

\cortext[cor1]{Corresponding author. Email: yuhao.kang@austin.utexas.edu}
\affiliation[doge]{organization={GISense Lab, Department of Geography and the Environment, The University of Texas at Austin},
            addressline={305 E 23rd St}, 
            city={Austin},
            postcode={78712}, 
            country={USA}}
\affiliation[doa]{organization={Department of Architecture, National University of Singapore},
            addressline={8 Architecture Drive}, 
            city={Singapore},
            postcode={117356}, 
            country={Singapore}}
\affiliation[dop]{organization={Department of Psychology, University of Texas at Austin},
            addressline={108 E Dean Keeton St}, 
            city={Austin},
            postcode={78712}, 
            country={USA}}
\affiliation[atof]{organization={Atof Inc.},
            addressline={17 Magazine St},
            city={Cambridge},
            postcode={02139},
            country={USA}}
\affiliation[dre]{organization={Department of Real Estate, National University of Singapore},
            addressline={15 Kent Ridge Drive}, 
            city={Singapore},
            postcode={119245}, 
            country={Singapore}}

\begin{abstract}
Human-environment interactions, a classic topic in geography, suggest that individuals and their environments might shape each other.
Yet the specific mechanisms underlying these interactions regarding human personality traits have not been explored. This study examines the associations between human Big Five personality traits and built environment characteristics derived from street view imagery across four cities in Texas, United States, providing a descriptive foundation for understanding these complex human-environment dynamics. By integrating fine-resolution self-reported personality assessments with computer vision analysis of urban environments, we identified significant spatial clustering of personality traits at the ZIP code level. Our regression analyses reveal that built environment features and socioeconomic characteristics explain substantial variance in personality distributions, with Openness showing the strongest model fit (R$^{2}$ = 0.47), followed by Agreeableness, Conscientiousness, Extraversion, and Neuroticism. Grouped built environment categories, socioeconomic factors, and demographic composition showed trait-specific patterns of association. These findings illustrate how personality traits may be associated with physical spaces at a smaller geographic scale than previously examined. Our results provide empirical evidence for understanding the link between psychological characteristics and environmental features, which can potentially enrich geography studies from a human-centered perspective.
\end{abstract}

\begin{graphicalabstract}
\includegraphics[width=\textwidth]{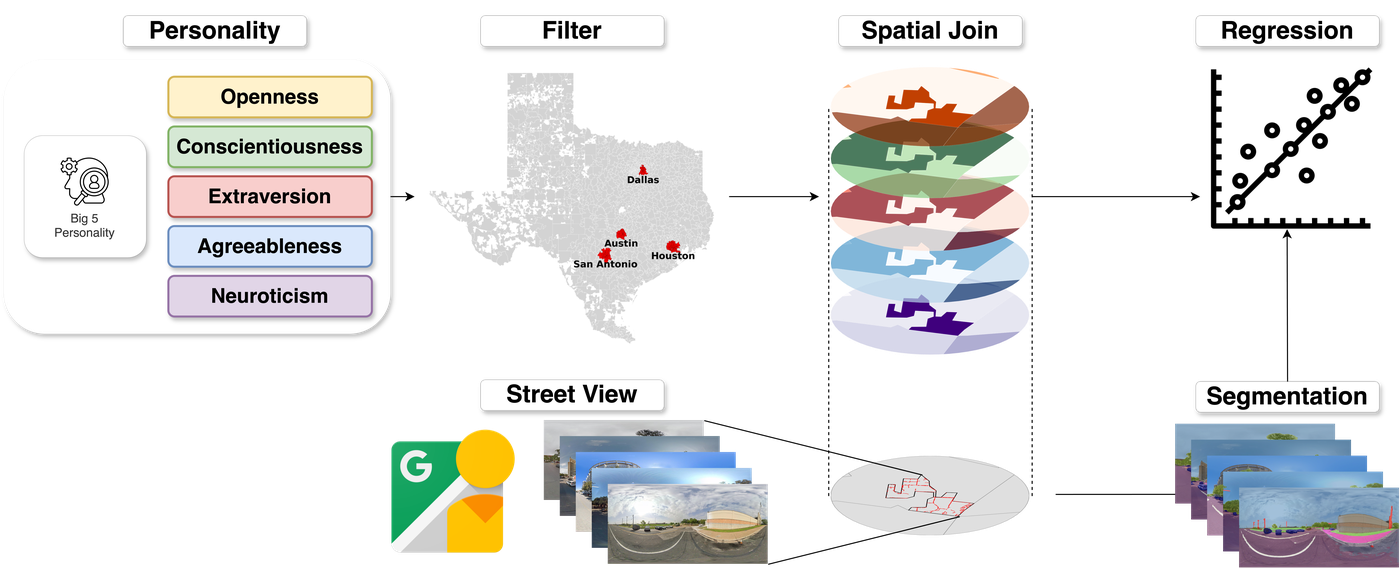}
\end{graphicalabstract}

\begin{highlights}
\item Big Five personality traits show spatial clustering at the neighborhood level
\item Personality traits have significant associations with built environment features
\item Study advances geographic psychology through GeoAI approaches and fine-scale analysis
\item Findings motivate future research to understand human-environment interactions
\end{highlights}


\end{frontmatter}


\section{Introduction}
\label{sec:intro} 

The built environment serves as the setting for most daily human experiences. 
Geography studies have long established that places significantly influence human perception, behaviors, health outcomes, and overall well-being through their physical and social characteristics \citep{rentfrow_geographical_2013, tuan_topophilia_1990, tuan_place_1975, gotz_unified_2025}.
Urban spaces vary considerably in their design, infrastructure, land use patterns, and amenities, creating distinct environments that may attract or influence different types of people. 
These unique place characteristics contribute to the formation of place identities and potentially shape or attract individuals with specific psychological profiles to particular locations \citep{jokela_are_2015, lalli_urbanrelated_1992, relph_place_1976, proshansky_placeidentity_1983, hidalgo_place_2001}.

Over the past decades, researchers have paid much attention to understanding human-environment relationships across various disciplines, including geography, urban planning, environmental psychology, and public health \citep{kabisch_human_2015, seymour_human_2016, steg_environmental_2018}.
Prior studies have examined how geographic and environmental features affect physical activity levels \citep{yang_examining_2023}, mobility patterns \citep{ito_assessing_2021}, perceived safety \citep{zhang_perception_2021}, and health outcomes \citep{zhang_utilizing_2024}. 
However, the human aspect of these relationships has been predominantly studied through behavioral outcomes or perceptual assessments, while more fundamental psychological characteristics have received less attention. 

Personality traits, referring to individuals' stable patterns of thinking, feeling, and behaving, have been relatively underexplored in the literature of geography, and there is untapped potential to uncover how people relate to and interact with their environments \citep{ebert_are_2022}. These traits collectively form what is referred to as personality: the enduring set of psychological characteristics that differentiate individuals from one another consistently across different situations \citep{bergner_what_2020}.
These distinct patterns of thinking, feeling, and behaving shape preferences, responses to stimuli, and interactions across various domains of life, including health \citep{leger_personality_2021}, social relationships \citep{back_personality_2023}, and work performance \citep{john_paradigm_2008, kang_associations_2023}.
The Big Five personality framework has been one of the most widely used measures of human personality traits in psychology \citep{gosling_very_2003, gurven_how_2013}. 
The five traits, consisting of Openness, Conscientiousness, Extraversion, Agreeableness, and Neuroticism, offer a widely used framework for understanding individual differences that has been widely validated across cultures and contexts \citep{john_paradigm_2008}.

Personality traits have demonstrated substantial geospatial variation across different regions, with diverse patterns emerging at national, regional, and even local scales \citep{jokela_are_2015, schmitt_geographic_2007, ebert_regional_2019, ebert_are_2022}.
These geographic variations suggest that geographic and environmental factors may play significant roles in shaping or attracting certain personality types to specific locations \citep{rentfrow_geographical_2020}. 
Previous research has identified associations between personality distributions and various geographical attributes, including climate \citep{wei_regional_2017}, topography \citep{gotz_physical_2020}, and regional socioeconomic characteristics \citep{rentfrow_geographical_2013}.
However, the specific elements of the built environment that might attract certain individuals or influence human personality traits have not yet been studied, particularly at finer neighborhood and street scales.

Built environments vary considerably in their spatial composition (e.g., the amount of greenery compared to that of buildings) \citep{oneill_indices_1988, he_aggregation_2000}. They may play a role in both shaping people's personalities and being shaped by them. For example, the physical characteristics of spaces may influence psychological development \citep{whipple_physical_2022}, while individuals with different personality profiles may selectively migrate to environments or may create urban planning norms that match their psychological preferences \citep{vandor_are_2021, amsalem_personality_2023}.

Despite the theoretical importance of understanding these human-environment interactions, detailed analyses examining how specific geographic and environmental elements relate to personality distributions have been limited by methodological challenges in quantifying environmental characteristics at scale. Theoretically, personality-environment relationships may operate through two interconnected processes: selective migration, where individuals choose locations aligned with their traits \citep{preston_personal_1981, kristof-brown_personorganization_2023, jokela_personality_2021, jokela_selective_2020}, and environmental influence, where built characteristics shape personality development through direct exposure and social interactions \citep{hopwood_genetic_2011, laceulle_transactions_2018, nieuwenhuis_neighbourhood_2021}. These processes likely operate bidirectionally, creating feedback loops where personality and place mutually shape each other \citep{matz_personality_2021}. By analyzing their associations, we can deepen our understanding of the mechanisms that link psychological traits to environmental features.

Recent technological advances in street view imagery (SVI) collection and artificial intelligence provide new opportunities to overcome these challenges. 
SVI has emerged as a rich data source for analyzing the built environment, providing detailed visual information about urban spaces from a pedestrian perspective \citep{biljecki_street_2021}. 
When combined with computer vision and machine learning techniques, this imagery enables researchers to quantify various elements of the urban landscape.
Prior studies have leveraged SVI to objectively and efficiently assess various built environmental characteristics, such as architectural styles \citep{liang_evaluating_2024}, green space coverage \citep{chen_quantifying_2020}, and pavement quality \citep{randhawa_paved_2024}. 
Recent studies using SVI have begun exploring relationships between perceived neighborhood characteristics and psychological variables, suggesting potential connections between visual environmental features and personality traits \citep{ito_understanding_2024, quintana_global_2025}.
However, few studies have examined how specific environmental characteristics might relate to personality traits, particularly on a large scale and across multiple cities.

This study aims to bridge existing research gaps by exploring the associations between the Big Five personality traits and built-environment characteristics derived from SVI across four cities in Texas. 
Compared to previous studies that have typically analyzed broader geographic units such as states or counties, our approach uses fine-scale data at the neighborhood (ZCTA) level, where individuals directly experience their environments.
For the first time in the literature, we analyze a dataset of Big Five personality traits alongside computer vision analysis of street-level imagery and investigate two primary research questions: \\
(1) What are the geographic patterns of human personality in the selected Texas cities?\\
(2) What are the associations between human personality traits and specific elements of the built environment?

The findings from this study have potential implications for geography, placemaking, public health, and community well-being. 
Understanding how different personality types interact with and respond to various environmental features may help planners create more psychologically supportive urban spaces \citep{chen_study_2024}. 
This paper advocates for integrating human-centered insights in geography by working with psychological datasets and perspectives \citep{kang_humancentered_2025} and potentially opens new opportunities to bring psychological perspectives into geography studies. 

\section{Literature Review}
\label{sec:litrev}

The exploration of relationships between human personality and built environment features requires a cross-disciplinary approach drawing from both geocomputing methodologies and psychological perspectives. This section reviews relevant literature across two domains: the analysis of built environments using SVI and geospatial artificial intelligence (GeoAI) \citep{biljecki_street_2021, zhang_urban_2024}, and research on personality across place from a psychological perspective \citep{rentfrow_geographical_2020}.

\subsection{Built Environment Analysis with Street View Imagery and AI}

SVI has emerged as a valuable data source for analyzing the built environment at unprecedented scales and resolutions. The availability of extensive image databases from providers such as Google Street View, combined with advances in computer vision and machine learning, has enabled a detailed assessment of urban landscapes that was previously infeasible through traditional field surveys \citep{biljecki_street_2021}. This methodology has transformed various domains of environmental analysis, from urban planning to public health research.

Prior studies using SVI have investigated multiple dimensions of the built environment. In urban form research, these images have been used to quantify land use patterns ~\citep{cao_integrating_2018}, green space distribution ~\citep{2023_jag_svi_sensitivity,chen_quantifying_2020}, and transportation infrastructure ~\citep{ito_assessing_2021}. Studies of the physical environment have employed SVI to assess environmental quality indicators such as air pollution \citep{oregan_associations_2022}, noise levels \citep{huang_estimating_2023}, and urban heat islands \citep{klimenka_instant_2025} through visual proxies. Social science research has explored relationships between visual elements of neighborhoods and phenomena like crime rates \citep{hipp_measuring_2022}, perceived safety \citep{zhang_perception_2021, zhou_multiscale_2024}, and social cohesion \citep{liu_natural_2020}. Human geography studies have examined humans' attention to street designs \citep{yang_urban_2024} and physical disorder of streets \citep{ma_measuring_2025}. Health-related research has utilized these images to characterize food environments \citep{tse_retrospective_2025}, investigate physical activity affordances \citep{yang_examining_2023}, and examine health outcomes \citep{zhang_utilizing_2024, kang_review_2020}.

There is also a large body of literature on assessing human subjective perceptions based on SVI.
One of the early studies is Place Pulse \citep{salesses_collaborative_2013, naik_streetscore_2014, dubey_deep_2016}, which collected human perception scores on safety, beauty, liveliness, wealthiness, boring-ness, and depressing-ness of street view images.
Other papers also looked into other perceptual characteristics of streets, such as building aesthetics \citep{liang_evaluating_2024}, playability \citep{kruse_places_2021}, and walkability \citep{kang_assessment_2023}. 

While there is a growing body of literature on SVI and human perceptions, these studies often focus primarily on perceptual aspects and tend to overlook other critical psychological dimensions. There is a notable gap in research regarding how individuals with diverse backgrounds and characteristics perceive, interact with, or are influenced by street characteristics \citep{kang_assessing_2023}. Such a research gap also includes the associations between street features and personality traits. Bridging this research gap matters because understanding these associations could clarify the bidirectional interactions between humans and their environments and potentially help geographers find interventions to foster urban environments to bring forth positive perceptual and psychological outcomes.

\subsection{Personality across Place}

The relationship between personality and geography has been rarely studied in geography. Despite this gap, psychologists have made efforts to examine spatial patterns of personality traits, though they have not typically examined detailed place characteristics or used advanced geocomputing methods.
Personality traits show spatial distributions and relationships with environmental factors, indicating that personality and place characteristics might shape each other \citep{ebert_are_2022}. These traits, as stable patterns of thinking, feeling, and behaving, are key constructs for examining person-environment interactions. Theoretical frameworks explain geographic variation in psychological traits through mechanisms like selective migration, environmental influence, social influence, and ecological influence \citep{rentfrow_theory_2008, mccrae_human_2004, wei_regional_2017}. These mechanisms underscore the bidirectional nature of human-environment relationships, forming the conceptual foundation of our study.

Empirical research has documented substantial regional differences in personality trait distributions across countries, states, and counties \citep{schmitt_geographic_2007, ebert_regional_2019, ebert_are_2022}. These patterns have been linked to various geographic factors. Historical factors such as environmental exposures (e.g., atmospheric lead levels) \citep{schwaba_impact_2021} and historical events (e.g., investment in infrastructure in the past and bombing during World War II) \citep{obschonka_did_2017, obschonka_roma_2025} have been examined as potential drivers of regional personality differences. Contemporary factors such as climate conditions, ambient temperatures \citep{wei_regional_2017}, land cover \citep{militaru_lay_2024}, and life satisfaction \citep{jokela_geographically_2015} have also shown associations with personality distributions.

Despite these advances, research on personality and geography has predominantly examined broad regional characteristics rather than specific features of the built environment. Most studies have used administrative boundaries (countries \citep{schmitt_geographic_2007}, states \citep{ebert_regional_2019}, counties \citep{ebert_are_2022}) as the geographic unit of analysis, potentially obscuring relationships that operate at finer neighborhood scales where individuals directly experience their environment and where the bidirectional interactions between people and places may be most pronounced. Additionally, the environmental characteristics examined by prior research have typically been macroscale factors such as climate or economic conditions rather than the physical design elements that constitute residents' immediate surroundings \citep{ben-shahar_real_2014, wei_regional_2017}. Most of these studies start from a psychological perspective, not from a geographic perspective.

Understanding geographic personality patterns is important because personality traits have been shown to relate to numerous societal outcomes at the individual level, including health behaviors \citep{leger_personality_2021, proto_covid19_2021}, economic decisions \citep{donnellan_personal_2009}, and social engagement \citep{wang_personality_2016}. This underscores the potential value of examining the associations between personality distributions and built environment characteristics at the fine-grain resolution, which could inform place-based interventions and policy decisions.

Failing to consider micro-scale built environment characteristics risks missing critical insights into the mechanisms underlying person–environment interactions. Addressing this gap from a geospatial perspective is essential, because it offers the opportunity to focus more on micro-level elements in the built environments by using GIS and advanced geocomputing, enabling a deeper understanding of how physical surroundings relate to psychological traits.
Understanding these relationships could have important implications for urban design, offering insights into how specific environmental modifications might support psychological well-being for different personality types. Furthermore, such research with GIS and GeoAI methods could inform our theoretical understanding of human-environment interactions by examining specific environmental elements most relevant to personality traits. 

Two theoretical mechanisms help explain the geographic variation in personality traits observed in prior research. Selection processes occur when individuals with particular personality profiles move toward environments that match their psychological preferences, or away from those that do not, as shown by work on personality's influence on aesthetic preferences for residential features \citep{dhumad_aesthetic_2024} and civic participation in urban planning \citep{li_personality_2025}. Because in-migration and out-migration may follow different dispersal patterns, the two directions need not produce symmetric geographic sorting. Selection can also operate indirectly: people often migrate for jobs, family, or housing costs, and these drivers are themselves correlated with physical features of the destination, so neighborhoods end up sorting residents on personality even when those physical characteristics are not the explicit motivation for moving. Environmental influence processes operate when built environment characteristics shape personality through direct exposure and social interactions \citep{hopwood_genetic_2011, laceulle_transactions_2018}. For example, daily exposure to walkable, socially active streets may reinforce extraverted and agreeable tendencies through repeated face-to-face encounters, while longer-term environmental exposures such as atmospheric lead have been linked to differences in adult personality \citep{schwaba_impact_2021}. Because selection and influence operate at the same time, they likely form a bidirectional feedback loop in which people shape the places they live and those places in turn shape the people who live there \citep{matz_personality_2021}. Empirical research has yet to systematically examine how specific built environment elements relate to these mechanisms at the neighborhood scale.

Our study addresses these limitations by examining associations between personality traits and specific built-environment features at the neighborhood level. By leveraging SVI analysis and comprehensive personality assessments, we bridge the methodological approaches of geocomputing and psychological perspectives to provide a descriptive foundation for understanding how human psychological characteristics relate to built environments—a necessary first step toward elucidating the complex mechanisms through which people and places shape each other.

\section{Data and Methods}
\label{sec:data}

\subsection{Conceptual Framework}
\begin{figure}[htbp]
\centering
\includegraphics[width=\textwidth]{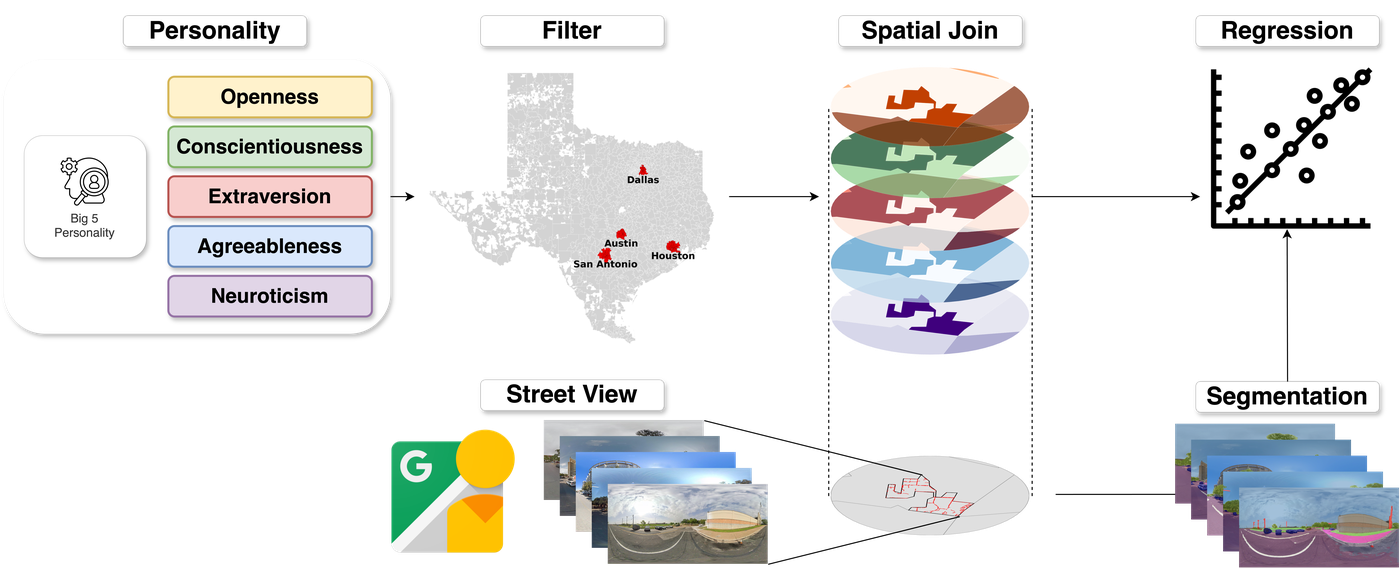}
\caption{Conceptual framework illustrating the data processing and analytical pipeline of this study.}
\label{fig:framework}
\end{figure}

This study employs an analytical framework to investigate the associations between human personality traits and the built environment. 
Figure \ref{fig:framework} illustrates our methodological approach, which integrates personality data with SVI analysis. 
Our analytical process involves several key steps: 
(1) data preparation;
(2) data processing (spatial join, image segmentation); and 
(3) correlation/regression analysis.

\subsection{Study Area Selection}
We selected four cities—Austin, Dallas, Houston, and San Antonio— as our study areas for several reasons. These cities represent the four largest urban centers in Texas, providing a comparable regional context in terms of governance structure, climate, biome, and shared challenges such as rapid urbanization and extensive sprawl. These similarities enhance the comparability of findings across locations. Meanwhile these cities also represent a diverse range of urban environments within Texas, each with unique demographic, cultural, and economic characteristics. This diversity allows for a comprehensive analysis of how personality traits interact with varied urban configurations, providing insights that are potentially generalizable to other urban contexts. Additionally, these cities have well-documented street view imagery, facilitating detailed environmental analysis through computer vision techniques. The selection of these cities thus ensures a robust examination of the associations between personality traits and built environment characteristics across different urban settings in Texas.

\subsection{Personality Dataset}
The personality data utilized in this study were obtained from the global Gosling-Potter Internet Project \citep{gosling_should_2004}, a large-scale online assessment platform that measures the Big Five personality traits: Openness, Conscientiousness, Extraversion, Agreeableness, and Neuroticism \citep{ebert_are_2022}. 
The platform employs the Big Five Inventory (BFI) and its successor the Big Five Inventory-2 (BFI-2), both of which are widely validated measures that recover the same Big Five trait structure closely enough for responses to be pooled across versions \citep{soto_next_2017}.
These traits represent a widely accepted framework in psychological research for understanding human personality differences. 
The dataset includes self-reported personality measures from participants who provided their residential ZIP code at the time of assessment, allowing for geographic aggregation and analysis. The personality data used in this study were collected between 2010 and 2020, while the SVI data span from 2007 to 2025. Although these collection windows overlap substantially (see \autoref{fig:temporal_histogram} in the Appendix), we acknowledge that temporal misalignment between the two data sources may introduce some noise, particularly in rapidly developing neighborhoods. Creating ZIP-code-by-year panel data would reduce sample sizes within each cell to levels insufficient for reliable estimation, so we aggregate responses across years.

The Big Five traits are operationalized as follows:
\begin{itemize}
    \item \textbf{Openness}: The degree of intellectual curiosity, creativity, and preference for novelty.
    \item \textbf{Conscientiousness}: The degree of organization, responsibility, and dependability.
    \item \textbf{Extraversion}: The degree of sociability, assertiveness, and positive emotionality.
    \item \textbf{Agreeableness}: The degree of altruism, cooperation, and empathy.
    \item \textbf{Neuroticism}: The degree of emotional instability, anxiety, and negative emotionality.
\end{itemize}

For our geographic analysis, we mapped participants' self-reported ZIP codes to Census ZIP Code Tabulation Areas (ZCTAs), which are generalized areal representations of USPS ZIP code service areas and serve as the spatial unit of analysis throughout this study. We aggregated individual responses at the ZCTA level, calculating mean scores for each trait after appropriate data cleaning and validation procedures.
At this stage, to ensure the robustness of our analysis, we filtered out observations with insufficient data (ZCTAs with fewer than 20 participants or less than 0.03\% coverage of 100 m grid points for Street View images).
This aggregation enables spatial analysis while maintaining participant privacy. Table \ref{tab:personality_summary} presents the summary statistics of the Big Five personality traits in our dataset.

\begin{table}[htbp]
\centering
\caption{Summary Statistics of Big Five Personality Traits at ZIP Code Level}
\label{tab:personality_summary}
\begin{tabular}{lrrrrr}
\toprule
\textbf{Trait} & \textbf{Mean} & \textbf{Median} & \textbf{Std. Dev.} & \textbf{Min} & \textbf{Max} \\
\midrule
Extraversion & 3.298 & 3.300 & 0.098 & 2.406 & 3.580 \\
Agreeableness & 3.775 & 3.772 & 0.086 & 3.317 & 4.074 \\
Conscientiousness & 3.603 & 3.599 & 0.109 & 3.033 & 4.148 \\
Neuroticism & 2.872 & 2.873 & 0.093 & 2.559 & 3.362 \\
Openness & 3.826 & 3.826 & 0.151 & 3.220 & 4.375 \\
Participants per ZIP & 442.371 & 352.500 & 523.896 & 3.000 & 7961.000 \\

\midrule
\multicolumn{6}{l}{\textbf{Dataset Characteristics}} \\
\midrule
Number of ZIP Codes & \multicolumn{5}{l}{340} \\
Total Participants & \multicolumn{5}{l}{150,406} \\

\bottomrule
\end{tabular}
\end{table}

\subsection{Street View Imagery and Computer Vision Analysis}
To capture detailed characteristics of the built environment, we utilized about five million Google Street View images collected on a 100-meter grid across the four selected Texan cities: Austin, Dallas, Houston, and San Antonio. 
This approach provided comprehensive visual coverage of the urban landscape, including all available historical imagery to maximize spatial coverage.

The street view images were processed using advanced computer vision techniques to extract quantitative measures of the built environment. 
Two primary analytical approaches were employed with a Python package called ZenSVI \citep{ito_zensvi_2025}:

\begin{itemize}
    \item \textbf{Semantic Segmentation}: This technique classifies each pixel in an image into predefined categories (e.g., buildings, roads, vegetation, sky), allowing for precise quantification of visual elements in the urban environment. We used the Mask2Former model pre-trained on the Mapillary Vistas dataset, which has been used by many urban studies \citep{cheng_maskedattention_2022, ito_examining_2024}.
    \item \textbf{Object Detection}: This approach identifies specific objects within images (e.g., American flags, automobiles, pedestrians), providing counts and spatial distributions of discrete elements in the urban landscape. We used GroundingDINO developed by \citet{liu2023grounding}, which is an object detection model that can detect objects in images based on text prompts. 
\end{itemize}

Table \ref{tab:segmentation_variables} lists the grouped built environment categories used in our analysis and their constituent segmentation and detection variables.
The extracted features were then aggregated at the ZCTA level to align with the personality data, creating a comprehensive set of built environment variables for subsequent analysis.

\subsection{Statistical Analysis}
Our analytical approach began with exploratory spatial data analysis to identify geographic patterns in personality traits. We conducted Moran's I tests (\autoref{tab:morans_i_results}) to assess spatial autocorrelation in the distribution of each trait, which confirmed significant spatial clustering patterns and justified further investigation of environmental correlates.

\begin{table}[htbp]
\centering
\caption{Moran's I Test Results for Big Five Personality Traits}
\begin{tabular}{lcc}
\toprule
Trait & Moran's I & p-value \\
\midrule
Openness & 0.1890 & 0.0010 \\
Conscientiousness & 0.1772 & 0.0010 \\
Extraversion & 0.1212 & 0.0020 \\
Agreeableness & 0.3127 & 0.0010 \\
Neuroticism & 0.1362 & 0.0020 \\
\bottomrule
\end{tabular}
\label{tab:morans_i_results}
\end{table}

Following confirmation of spatial patterns, we developed Ordinary Least Squares (OLS) regression models and Spatial Lag Models (SLM) to examine associations between personality traits and built environment characteristics. 

For each of the Big Five traits, we specified the following OLS model:

\begin{equation}
P_i = \beta_0 + \sum_{j=1}^{n} \beta_j X_{ij} + \varepsilon_i
\end{equation}

Where $P_i$ represents the mean standardized score of a personality trait in ZCTA $i$, $\beta_0$ is the intercept, $X_{ij}$ represents the $j$th predictor variable in ZCTA $i$, $\beta_j$ is the corresponding coefficient, and $\varepsilon_i$ is the error term.

While the OLS model assumes that each observation is independent, spatial data often violate this assumption due to geographic proximity, as neighboring areas may exhibit similar personality patterns.
To account for this spatial dependence, we estimated the Spatial Lag Model (SLM), expressed as:

\begin{equation}
P_i = \rho \sum_{k} w_{ik} P_k + \sum_{j=1}^{n} \beta_j X_{ij} + \varepsilon_i
\end{equation}

Where $w_{ik}$ denotes the spatial weight between ZCTAs $i$ and $k$, defined by a spatial-weights matrix $W$ constructed using Queen contiguity with row-standardized weights. Under this specification, two ZCTAs are considered neighbors if they share a boundary or vertex, which captures the spatial adjacency structure of the study area.
The parameter $\rho$ represents the spatial autoregressive coefficient, quantifying the strength and direction of spatial dependence.
$X_{ij}$ represents the same set of predictor variables used in the OLS model, and $\varepsilon_i$ is the error term.

To improve model interpretability and reduce the number of predictors, we grouped the fine-grained segmentation and detection variables listed in Table \ref{tab:segmentation_variables} into eight theoretically informed categories based on established frameworks in environmental psychology and urban planning: (1) Greenery, (2) Open Space, (3) Building, (4) Road, (5) Active Mobility Infrastructure, (6) Active Mobility Presence, (7) Vehicle Presence, and (8) Physical Boundaries. For each category, we created composite scores by summing constituent variables (ratio variables or count variables, as appropriate), and the resulting composite scores were re-standardized before inclusion in the regression models. Three additional features were retained as individual predictors: Symbolic US Flag (American flag count), CCTV Surveillance (detected via object detection), and Visual Complexity (computed from semantic segmentation pixel ratios).

In addition to these environmental variables, we incorporated eight socio-demographic and socioeconomic variables and three city dummy variables to address confounding from neighborhood composition and to absorb city-level unobserved heterogeneity. The socio-demographic and socioeconomic variables were derived from the U.S. Census Bureau’s American Community Survey (ACS) 2020 5-Year Estimates at the ZCTA level \citep{USCensus2020ACS}. These include male share, four age group shares (15--29, 30--44, 45--59, and 60+), population density, median household income, and a racial diversity index. The three city dummy variables (Austin, Dallas, and Houston) were introduced to account for unobserved city-level heterogeneity, with San Antonio serving as the reference category. These binary variables function as city-fixed effects.

Prior to modeling, we performed several data processing steps to ensure statistical validity. Personality trait scores and environmental variables were standardized using Z-score transformation to facilitate the interpretation and comparison of effect sizes across different variables and cities, except for cities dummy variables, which accounts for city-specific fixed effects and unobserved heterogeneity. For environmental variables with significant skewness, we applied Box-Cox transformations to approximate normal distributions \citep{box_analysis_1964}. This transformation is particularly effective for addressing the positively skewed distributions common in count data from object detection (e.g., car counts, pedestrian counts).

Our variable selection process was based on bivariate correlation tests between built environment elements and personality traits. Only variables with significant correlations with at least one personality trait were retained for further analysis. We then conducted an iterative variance inflation factor (VIF) analysis to mitigate multicollinearity: at each step, we calculated VIF values for all remaining variables, removed the variable with the highest VIF value, and repeated until all remaining variables had VIF values below 10.

\section{Results}
\label{sec:results}

\subsection{Geographic Patterns of Personality Traits}
\begin{figure}[htbp]
\centering
\includegraphics[width=\textwidth]{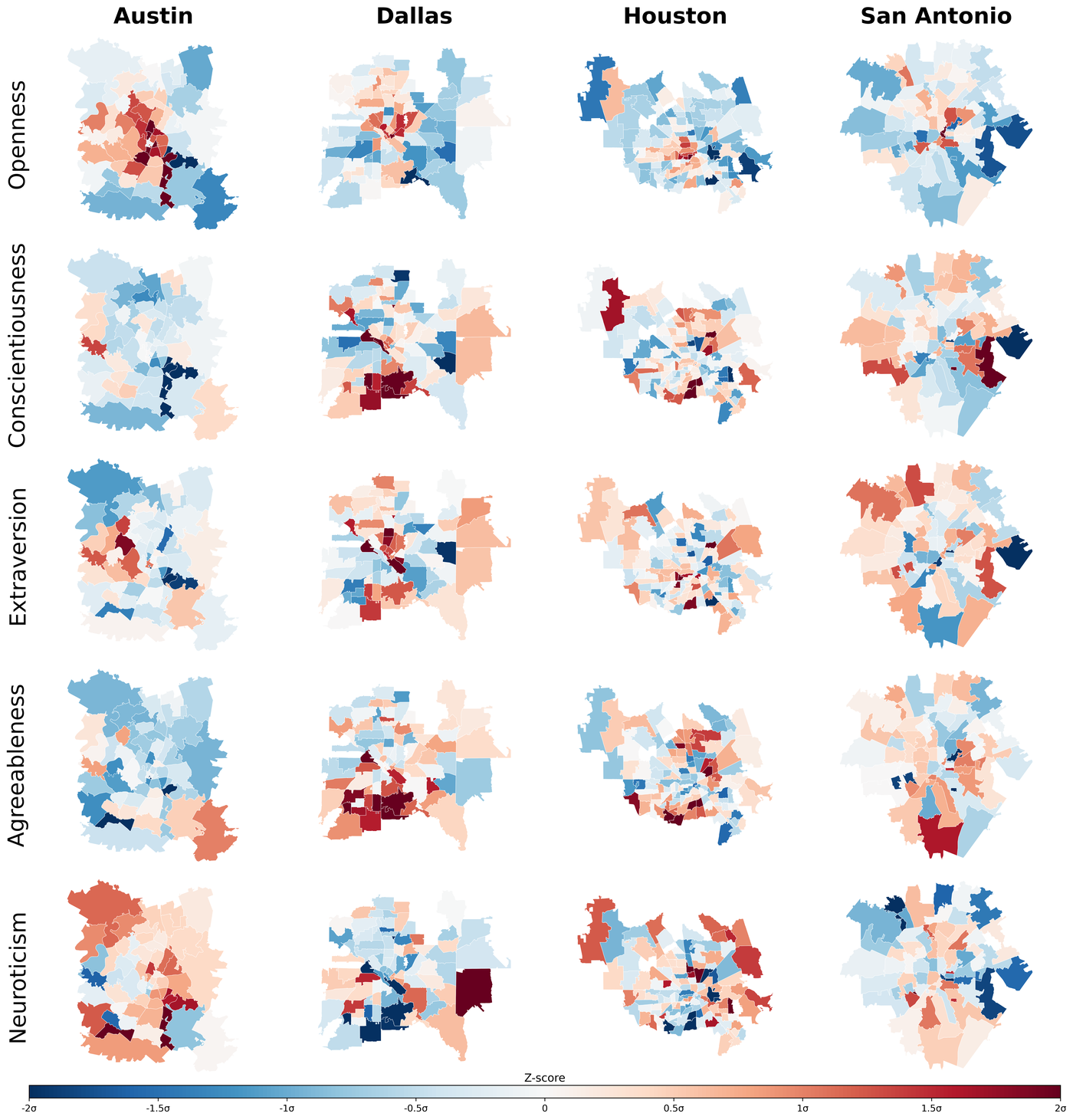}
\caption{Spatial distribution of Big Five personality traits across ZCTAs in Austin, Dallas, Houston, and San Antonio. Values are standardized as Z-scores, with red indicating above-average trait levels and blue indicating below-average levels relative to each city's mean.}
\label{fig:personality_maps}
\end{figure}

Analysis of the spatial distribution of personality traits across the four Texan cities revealed distinct geographic patterns with significant spatial autocorrelation. \autoref{fig:personality_maps} illustrates the spatial distribution of the Big Five personality traits -- Openness, Conscientiousness, Extraversion, Agreeableness, and Neuroticism -- across ZCTAs in Austin, Dallas, Houston, and San Antonio.

The visualized personality distributions exhibit intra-city variation, with some traits showing stronger spatial clustering than others. 
Openness shows clear spatial patterns with clusters of high values in the central neighborhoods of Austin and lower values in the peripheries, such as the southeastern neighborhoods of Austin, the western neighborhoods of Dallas, and the southern neighborhoods of San Antonio.
Conscientiousness demonstrates notable concentrations as well, particularly high in the southwestern neighborhoods of Dallas, whereas Austin exhibits clusters of lower Conscientiousness primarily in its southeastern neighborhoods. 
Extraversion shows high values centrally in Austin and Dallas, as well as the northeastern neighborhoods in San Antonio. Agreeableness shows low values for most of the Austin neighborhoods, while Southern Dallas and Southern Houston show high values. 
Neuroticism is high for most Austin neighborhoods and low in the southeastern Dallas areas, indicating pronounced local disparities. 
Among different personality traits, Conscientiousness and Agreeableness seem to have similar spatial distribution, whereas Neuroticism seems to be negatively correlated with them.
Collectively, these variations highlight nuanced local geographies of personality within urban environments, suggesting that there might be neighborhood-level characteristics associated with the personality traits.

Moran's I tests (\autoref{tab:morans_i_results}) confirm statistically significant spatial autocorrelation for all five personality traits, with Agreeableness showing the strongest spatial clustering (I = 0.3127, p < 0.001), followed by Openness (I = 0.1890, p < 0.001), Conscientiousness (I = 0.1772, p < 0.001), Neuroticism (I = 0.1362, p < 0.001), and Extraversion (I = 0.1212, p < 0.01). These results indicate that similar personality trait levels tend to cluster in geographically proximate areas, suggesting that neighborhood-level socio-demographic and environmental contexts may play a role in shaping or attracting specific personality types. This spatial autocorrelation warrants our investigation into potential environmental factors that might explain these geographic patterns in personality traits.

\subsection{Built Environment Characteristics}
\begin{figure}[htbp]
\centering
\includegraphics[width=\textwidth]{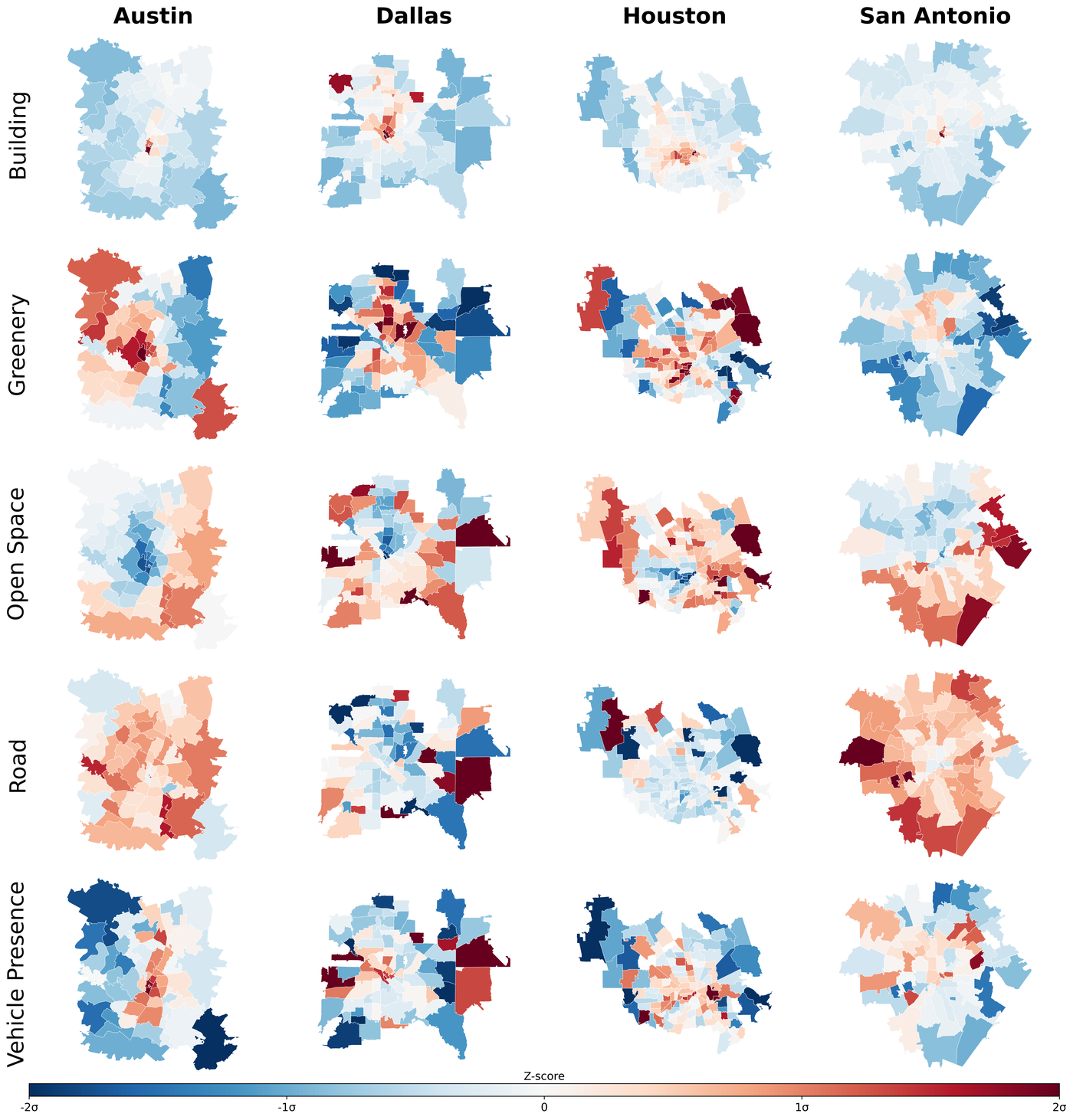}
\caption{Spatial distribution of five urban environmental features across ZCTAs in Austin, Dallas, Houston, and San Antonio. Values are standardized as Z-scores, with red indicating above-average feature levels and blue indicating below-average levels relative to each city's mean.}
\label{fig:environment_maps}
\end{figure}

The spatial distribution of key built environment features also exhibits distinct geographic patterns across the four cities, as illustrated in \autoref{fig:environment_maps}.
These maps reveal clear spatial heterogeneity in built environment features both across and within cities. 
Buildings have very high values in the central neighborhoods in all four cities, with a few exceptions in the northern area in Dallas. 
Car counts are as strongly clustered as buildings but are still consistently higher in central neighborhoods in all four cities, reflecting car-oriented urban layouts in these areas. 
Many neighborhoods in Dallas have higher values for fence compared to the other three cities.
Road markings have higher values in the peripheral neighborhoods across the four cities, and Austin has particularly higher values throughout the city.
Vegetation cover shows notable contrasts as well, with higher values in the central areas, with a few exceptions in the northwestern and southeastern neighborhoods in Austin, and a few peripheral neighborhoods in Houston.
These patterns underscore the spatial clustering of built environment features and mobility infrastructure, providing a foundation for examining their associations with neighborhood-level psychological traits.

\begin{figure}[htbp]
\centering
\includegraphics[width=0.75\textwidth]{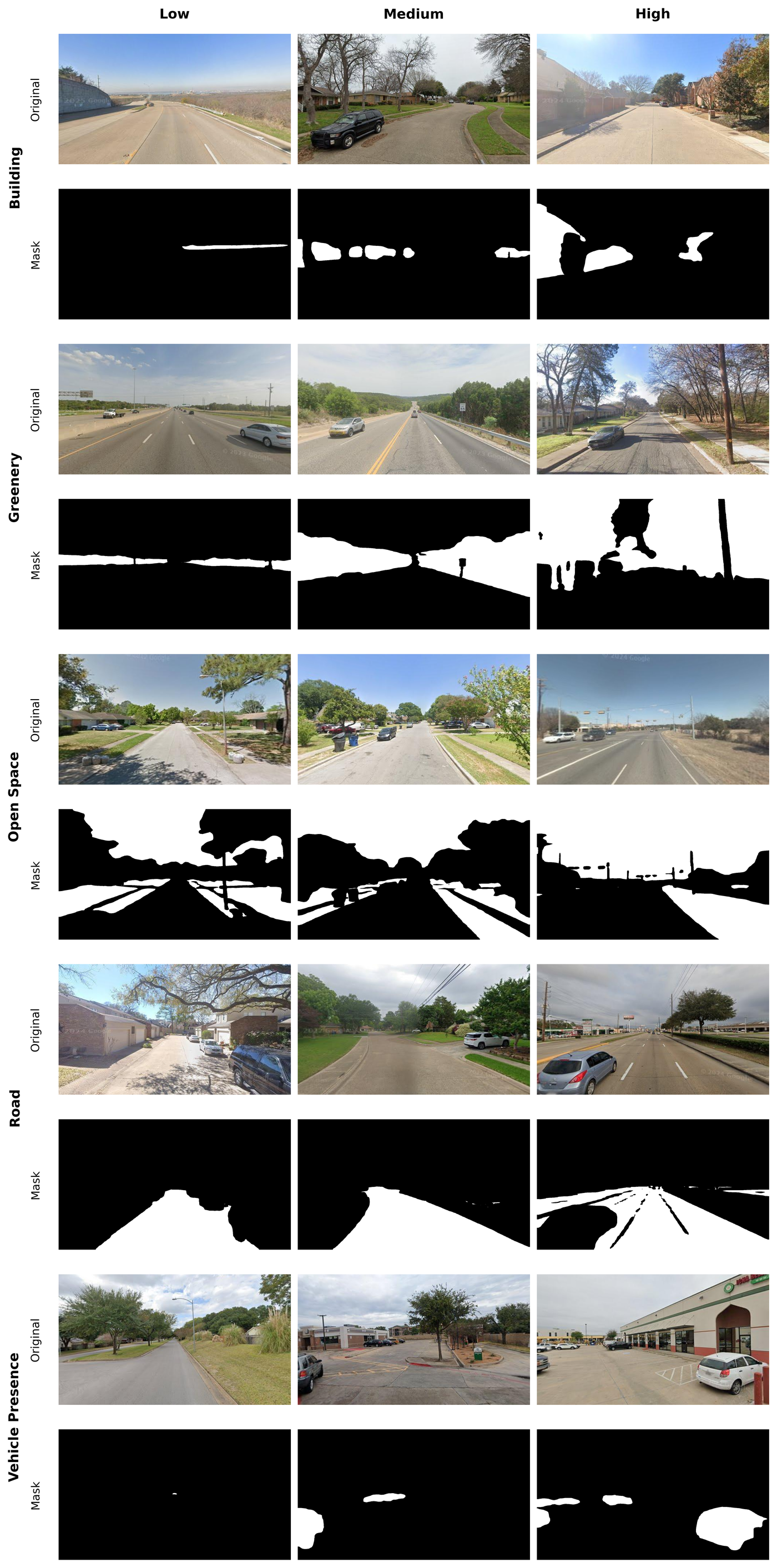}
\caption{Visual examples of built environment features across different intensity levels (high, medium, low) captured through SVI in the four study cities. A pair of rows (original and segmented) represents a different environmental feature (Building, Greenery, Open Space, Road, and Vehicle Presence), showing how these elements manifest across varying urban and peri-urban contexts.}
\label{fig:feature_grid}
\end{figure}

\autoref{fig:feature_grid} provides visual examples of how different built environment features manifest at high, medium, and low intensity levels across the study cities. The figure specifically illustrates five grouped categories: building, greenery, open space, road, and vehicle presence. These original and segmented images demonstrate the significant visual variation in environmental elements, from dense urban cores with high building density to suburban natural landscapes with minimal built infrastructure.

\subsection{Correlation Between Personality Traits and Built Environment Features}
\begin{figure}[htbp]
\centering
\includegraphics[width=0.75\textwidth]{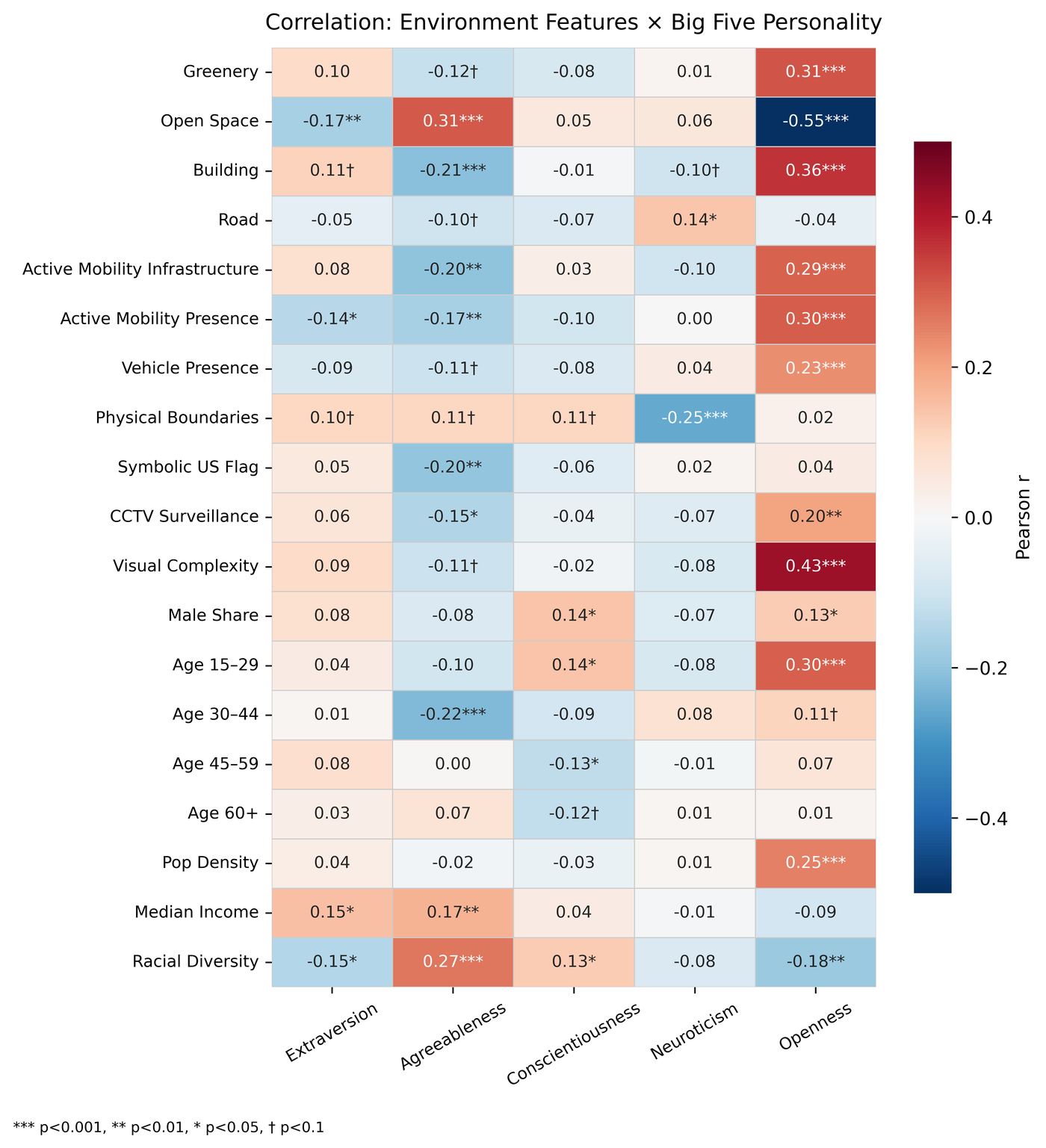}
\caption{Heatmap of Pearson correlation coefficients between Big Five personality traits and environmental and socioeconomic features. Color intensity indicates strength and direction of correlation, with significance levels marked by symbols ($\dagger$ p $<$ 0.1, * p $<$ 0.05, ** p $<$ 0.01, *** p $<$ 0.001).}
\label{fig:correlation_heatmap}
\end{figure}

Before conducting regression analyses, we examined bivariate correlations between personality traits and the grouped built environment features, additional SVI-derived features, and socioeconomic variables to identify potential associations. \autoref{fig:correlation_heatmap} displays the Pearson correlation coefficients between the Big Five personality traits and these predictor variables.

The correlation analysis revealed distinct patterns across personality traits. Openness showed the strongest and most numerous correlations, exhibiting strong positive associations with visual complexity (r = 0.43, p $<$ 0.001), building (r = 0.36, p $<$ 0.001), greenery (r = 0.31, p $<$ 0.001), active mobility presence (r = 0.30, p $<$ 0.001), active mobility infrastructure (r = 0.29, p $<$ 0.001), population density (r = 0.25, p $<$ 0.001), and vehicle presence (r = 0.23, p $<$ 0.001). Conversely, Openness showed a strong negative correlation with open space (r = $-$0.55, p $<$ 0.001) and a smaller negative association with racial diversity (r = $-$0.18, p $<$ 0.01). This pattern suggests that Openness is higher in visually complex, densely built urban environments and lower in areas dominated by open landscapes.

Agreeableness displayed a largely opposite pattern to Openness. It was positively associated with open space (r = 0.31, p $<$ 0.001), racial diversity (r = 0.27, p $<$ 0.001), and median income (r = 0.17, p $<$ 0.01), while showing negative correlations with building (r = $-$0.21, p $<$ 0.001), the share of residents aged 30--44 (r = $-$0.22, p $<$ 0.001), active mobility infrastructure (r = $-$0.20, p $<$ 0.01), and symbolic US flag (r = $-$0.20, p $<$ 0.01). These associations suggest that Agreeableness tends to be higher in more open, affluent, and racially diverse neighborhoods.

Conscientiousness showed few and weak correlations with built environment features; its most notable associations were instead with demographic composition, namely male share (r = 0.14, p $<$ 0.05) and the share of residents aged 15--29 (r = 0.14, p $<$ 0.05). Extraversion was negatively correlated with open space (r = $-$0.17, p $<$ 0.01), active mobility presence (r = $-$0.14, p $<$ 0.05), and racial diversity (r = $-$0.15, p $<$ 0.05), while positively associated with median income (r = 0.15, p $<$ 0.05). Neuroticism showed a notable negative correlation with physical boundaries (r = $-$0.25, p $<$ 0.001) and a positive association with road (r = 0.14, p $<$ 0.05).

These correlation patterns informed our variable selection and provided initial evidence of systematic associations between built environment characteristics and personality trait distributions across neighborhoods.

\subsection{Associations Between Personality Traits and Built Environment Features}

To investigate the relationships between personality traits and built environment characteristics, we conducted OLS and SLM regression analyses for each of the Big Five personality traits. As described in Section 3.4, the fine-grained environmental features were grouped into eight theoretically informed categories (Greenery, Open Space, Building, Road, Active Mobility Infrastructure, Active Mobility Presence, Vehicle Presence, and Physical Boundaries), along with three individual features (Symbolic US Flag, CCTV Surveillance, and Visual Complexity), as well as demographic and socioeconomic controls and city fixed effects. In total, 255 ZCTAs across the four study cities were retained for analysis (see \autoref{fig:zipcode_inclusion} in the Appendix for the spatial distribution of included and excluded ZIP codes).

Our OLS regression analysis results are shown in Table~\ref{tab:ols_betas_ours_v2}. The results indicate varying explanatory power across trait models, with the highest R$^{2}$ observed for Openness (R$^{2}$ = 0.467, adjusted R$^{2}$ = 0.416), followed by Agreeableness (R$^{2}$ = 0.353, adjusted R$^{2}$ = 0.292), Conscientiousness (R$^{2}$ = 0.211, adjusted R$^{2}$ = 0.136), Extraversion (R$^{2}$ = 0.180, adjusted R$^{2}$ = 0.102), and Neuroticism (R$^{2}$ = 0.157, adjusted R$^{2}$ = 0.077).

\begin{table}[htbp]
\centering
\footnotesize
\caption{OLS standardized coefficients ($\beta$) for Big Five personality traits and environmental predictors (pooled model, HC1 robust SE).}
\label{tab:ols_betas_ours_v2}
\resizebox{\textwidth}{!}{%
\begin{tabular}{lccccc}
\toprule
\textbf{Predictor} & \textbf{Extraversion} & \textbf{Agreeableness} & \textbf{Conscientiousness} & \textbf{Neuroticism} & \textbf{Openness} \\
\midrule
\multicolumn{6}{l}{\textit{Built environment features}} \\
Greenery & $-0.021$ & $-0.293$ & $-0.532^{**}$ & $0.177$ & $0.152$ \\
Open Space & $-0.161$ & $0.012$ & $-0.544^{**}$ & $0.200$ & $-0.128$ \\
Building & $0.286$ & $-0.295$ & $-0.370^{\dagger}$ & $0.035$ & $0.090$ \\
Road & $0.272$ & $-0.079$ & $-0.328^{\dagger}$ & $-0.055$ & $0.012$ \\
Active Mobility Infrastructure & $-0.113$ & $-0.097$ & $0.012$ & $0.043$ & $-0.039$ \\
Active Mobility Presence & $-0.144$ & $0.014$ & $-0.009$ & $-0.113$ & $0.047$ \\
Vehicle Presence & $-0.145^{\dagger}$ & $-0.034$ & $-0.109$ & $0.125$ & $0.036$ \\
Physical Boundaries & $0.030$ & $-0.262^{*}$ & $-0.297^{*}$ & $-0.124$ & $-0.090$ \\
\midrule
\multicolumn{6}{l}{\textit{Individual features and controls}} \\
Symbolic US Flag & $-0.049$ & $-0.130^{\dagger}$ & $-0.126^{\dagger}$ & $0.068$ & $-0.005$ \\
CCTV Surveillance & $-0.112$ & $0.094$ & $0.002$ & $-0.022$ & $-0.119$ \\
Visual Complexity & $0.093$ & $0.293^{\dagger}$ & $0.089$ & $-0.150$ & $0.080$ \\
Male Share & $-0.040$ & $-0.064$ & $0.117$ & $0.007$ & $-0.042$ \\
Age 15--29 & $0.221$ & $0.017$ & $0.188$ & $-0.216$ & $0.775^{***}$ \\
Age 30--44 & $-0.001$ & $-0.191^{\dagger}$ & $-0.054$ & $-0.015$ & $0.441^{***}$ \\
Age 45--59 & $0.064$ & $0.033$ & $-0.099$ & $-0.108$ & $0.307^{***}$ \\
Age 60+ & $-0.016$ & $0.083$ & $0.050$ & $-0.002$ & $0.356^{**}$ \\
Pop Density & $-0.111$ & $-0.075$ & $-0.210^{*}$ & $0.280^{**}$ & $-0.059$ \\
Median Income & $0.374^{**}$ & $0.379^{**}$ & $0.153$ & $-0.223$ & $0.003$ \\
Racial Diversity & $-0.170^{*}$ & $0.223^{**}$ & $0.014$ & $-0.094$ & $0.027$ \\
\midrule
$R^2$ & $0.180$ & $0.353$ & $0.211$ & $0.157$ & $0.467$ \\
\textit{Adj.}\ $R^2$ & $0.102$ & $0.292$ & $0.136$ & $0.077$ & $0.416$ \\
\bottomrule
\multicolumn{6}{l}{\scriptsize $^{***}p<0.001$, $^{**}p<0.01$, $^{*}p<0.05$, $^{\dagger}p<0.1$. OLS pooled model with city fixed effects} \\
\multicolumn{6}{l}{\scriptsize (San Antonio = baseline). All continuous predictors z-scored.} \\
\end{tabular}%
}
\end{table}

Openness showed the strongest overall model fit among all traits, though this was driven predominantly by demographic composition rather than built environment features. All four age group shares were highly significant: residents aged 15--29 ($\beta = 0.775$, $p < 0.001$), aged 30--44 ($\beta = 0.441$, $p < 0.001$), aged 45--59 ($\beta = 0.307$, $p < 0.001$), and aged 60 and above ($\beta = 0.356$, $p < 0.01$). No grouped built environment category reached conventional significance levels for Openness, suggesting that the spatial variation in Openness is more closely linked to neighborhood demographic composition than to physical environmental characteristics at this scale of analysis.

Conscientiousness exhibited the most extensive associations with built environment features among all traits. Significant negative associations were found with open space ($\beta = -0.544$, $p < 0.01$), greenery ($\beta = -0.532$, $p < 0.01$), building ($\beta = -0.370$, $p < 0.1$), road ($\beta = -0.328$, $p < 0.1$), physical boundaries ($\beta = -0.297$, $p < 0.05$), and symbolic US flag ($\beta = -0.126$, $p < 0.1$). Population density was also negatively associated ($\beta = -0.210$, $p < 0.05$). These results suggest that areas with more visible environmental features---including open landscapes, greenery, buildings, and road infrastructure---correspond to lower Conscientiousness, whereas less visually complex and less densely developed areas are associated with higher Conscientiousness levels.

Agreeableness had the second-highest model fit (R$^{2}$ = 0.353) with a distinct pattern of associations primarily shaped by socioeconomic factors and select built environment features. Median household income ($\beta = 0.379$, $p < 0.01$) and racial diversity ($\beta = 0.223$, $p < 0.01$) were positively associated, while physical boundaries ($\beta = -0.262$, $p < 0.05$) and symbolic US flag ($\beta = -0.130$, $p < 0.1$) showed negative associations. Visual complexity was positively associated at marginal significance ($\beta = 0.293$, $p < 0.1$), and the share of residents aged 30--44 showed a marginal negative association ($\beta = -0.191$, $p < 0.1$). These findings indicate that more affluent and racially diverse neighborhoods with fewer physical barriers and symbolic markers tend to exhibit higher Agreeableness.

Extraversion was primarily associated with socioeconomic characteristics. Median household income showed a positive association ($\beta = 0.374$, $p < 0.01$), while racial diversity was negatively associated ($\beta = -0.170$, $p < 0.05$). Among built environment features, only vehicle presence showed a marginal negative association ($\beta = -0.145$, $p < 0.1$), suggesting that areas with greater vehicle activity correspond to slightly lower Extraversion levels.

Neuroticism had the lowest model fit (R$^{2}$ = 0.157) and showed limited significant associations. Population density was positively associated ($\beta = 0.280$, $p < 0.01$), indicating that more densely populated areas tend to exhibit higher Neuroticism levels. No grouped built environment category or socioeconomic variable reached conventional significance for Neuroticism in the OLS model.

Our Spatial Lag Model (SLM) regression was conducted using the same set of predictors and observations as the OLS models to ensure comparability. The results are shown in Table~\ref{tab:sar_betas_ours_v2}. The spatial weights matrix was defined using Queen contiguity with row-standardized weights.

\begin{table}[htbp]
\centering
\footnotesize
\caption{Spatial lag standardized coefficients ($\beta$) for Big Five personality traits and environmental predictors (pooled model, Queen contiguity weights).}
\label{tab:sar_betas_ours_v2}
\resizebox{\textwidth}{!}{%
\begin{tabular}{lccccc}
\toprule
\textbf{Predictor} & \textbf{Extraversion} & \textbf{Agreeableness} & \textbf{Conscientiousness} & \textbf{Neuroticism} & \textbf{Openness} \\
\midrule
\multicolumn{6}{l}{\textit{Built environment features}} \\
Greenery & $-0.016$ & $-0.270^{\dagger}$ & $-0.506^{**}$ & $0.178$ & $0.154$ \\
Open Space & $-0.149$ & $0.012$ & $-0.510^{**}$ & $0.202$ & $-0.089$ \\
Building & $0.324$ & $-0.286^{\dagger}$ & $-0.305$ & $0.050$ & $0.090$ \\
Road & $0.282^{\dagger}$ & $-0.045$ & $-0.323^{*}$ & $-0.035$ & $0.008$ \\
Active Mobility Infrastructure & $-0.124$ & $-0.083$ & $-0.001$ & $0.046$ & $-0.038$ \\
Active Mobility Presence & $-0.146$ & $-0.019$ & $-0.051$ & $-0.107$ & $0.039$ \\
Vehicle Presence & $-0.165^{\dagger}$ & $-0.024$ & $-0.108$ & $0.108$ & $0.026$ \\
Physical Boundaries & $0.010$ & $-0.217^{*}$ & $-0.290^{*}$ & $-0.130$ & $-0.106$ \\
\midrule
\multicolumn{6}{l}{\textit{Individual features and controls}} \\
Symbolic US Flag & $-0.030$ & $-0.135^{*}$ & $-0.109$ & $0.080$ & $-0.005$ \\
CCTV Surveillance & $-0.148$ & $0.100$ & $-0.042$ & $-0.040$ & $-0.110$ \\
Visual Complexity & $0.145$ & $0.278^{\dagger}$ & $0.140$ & $-0.113$ & $0.071$ \\
Male Share & $-0.033$ & $-0.050$ & $0.135^{\dagger}$ & $0.009$ & $-0.045$ \\
Age 15--29 & $0.200$ & $0.061$ & $0.153$ & $-0.210$ & $0.717^{***}$ \\
Age 30--44 & $-0.013$ & $-0.155^{\dagger}$ & $-0.064$ & $-0.013$ & $0.431^{***}$ \\
Age 45--59 & $0.035$ & $0.070$ & $-0.114$ & $-0.118$ & $0.297^{***}$ \\
Age 60+ & $-0.016$ & $0.101$ & $0.025$ & $0.011$ & $0.325^{***}$ \\
Pop Density & $-0.129$ & $-0.038$ & $-0.209^{*}$ & $0.271^{**}$ & $-0.066$ \\
Median Income & $0.393^{*}$ & $0.328^{*}$ & $0.116$ & $-0.208$ & $-0.013$ \\
Racial Diversity & $-0.153^{\dagger}$ & $0.197^{**}$ & $0.018$ & $-0.083$ & $0.036$ \\
\midrule
Pseudo $R^2$ & $0.187$ & $0.404$ & $0.252$ & $0.163$ & $0.479$ \\
\bottomrule
\multicolumn{6}{l}{\scriptsize $^{***}p<0.001$, $^{**}p<0.01$, $^{*}p<0.05$, $^{\dagger}p<0.1$. Spatial lag model (ML\textsubscript{Lag}), pooled with city FE} \\
\multicolumn{6}{l}{\scriptsize (San Antonio = baseline), Queen contiguity, row-standardized weights. All continuous predictors z-scored.} \\
\end{tabular}%
}
\end{table}

Comparing the Pseudo R$^{2}$ values from SLM with the R$^{2}$ values from the OLS models, all five traits showed improvement after accounting for spatial dependence. The most notable gains were for Agreeableness (from 0.353 to 0.404) and Conscientiousness (from 0.211 to 0.252), suggesting that spatial spillover effects are particularly relevant for these traits. Openness increased from 0.467 to 0.479, Extraversion from 0.180 to 0.187, and Neuroticism from 0.157 to 0.163.

When comparing the two models, most coefficient estimates remained stable in direction and magnitude. For Conscientiousness, the SLM strengthened the significance of road ($\beta = -0.323$, $p < 0.05$) and introduced a marginally significant positive association with male share ($\beta = 0.135$, $p < 0.1$). For Agreeableness, several additional built environment features gained significance in the spatial model, including greenery ($\beta = -0.270$, $p < 0.1$), building ($\beta = -0.286$, $p < 0.1$), and symbolic US flag ($\beta = -0.135$, $p < 0.05$). For Extraversion, road gained marginal significance ($\beta = 0.282$, $p < 0.1$). Openness and Neuroticism results remained largely unchanged from the OLS estimates.

Across traits, the grouped built environment categories revealed interpretable patterns. Although Openness exhibited the strongest and most numerous bivariate correlations with built environment features, Conscientiousness emerged as the trait most sensitive to those features once demographic and socioeconomic controls were included, with five of eight grouped categories and symbolic US flag reaching at least marginal significance in the OLS specification, all in the negative direction. This consistent pattern suggests that higher environmental complexity and infrastructure density correspond to lower Conscientiousness levels. Agreeableness was shaped by a combination of socioeconomic factors (income, racial diversity) and specific built environment features (physical boundaries, symbolic US flag), with additional associations with greenery and building emerging in the spatial model. Openness, despite having the highest overall R$^{2}$, was driven predominantly by demographic age composition rather than built environment features. Extraversion and Neuroticism showed relatively few significant associations with built environment categories, with socioeconomic and density factors playing more prominent roles. These findings support theoretical frameworks suggesting that physical environments may be associated with personality trait distributions, whether through shaping personality development or attracting individuals with specific personality profiles to particular neighborhood types \citep{rentfrow_theory_2008}.

\section{Discussion}
\label{sec:discussion}

Our findings reveal meaningful associations between built environment features derived from SVI and Big Five personality traits, providing new evidence on how urban spaces relate to personality traits. We emphasize that these relationships are correlational in nature, and the observed associations likely reflect complex underlying human-environment interaction processes. Many of these correlations could be driven by third variables, such as historical development patterns or social environments that influence both personality distributions and built environment characteristics. For instance, features like transportation infrastructure may not directly attract or repel individuals with certain personality traits, but rather serve as proxies for broader neighborhood characteristics that affect residential choices. With this important caveat in mind, we discuss the theoretical mechanisms and broader implications of these findings.

\subsection{Interpreting Personality-Environment Associations}

As outlined in Section \ref{sec:intro}, the associations we observe can be understood through two plausible processes that may operate simultaneously: selective migration (personality shapes location choice) and environmental influence (built environment shapes personality). While our cross-sectional design precludes causal inference, examining our trait-specific results through this theoretical lens yields interpretable patterns.

Conscientiousness exhibited the most extensive built-environment associations, with five of eight grouped categories and symbolic US flag showing at least marginal significance in the OLS model, all in the negative direction. This consistent pattern—spanning greenery, open space, building, road, physical boundaries, and symbolic US flag—suggests that areas with more prominent and varied environmental features correspond to lower Conscientiousness levels. From a selection perspective, individuals high in Conscientiousness, who tend to prefer structured and orderly environments \citep{john_paradigm_2008}, may gravitate toward less visually complex neighborhoods. From an environmental influence perspective, less stimulating physical surroundings may reinforce traits associated with self-discipline and order \citep{hopwood_genetic_2011}.

Agreeableness was primarily shaped by socioeconomic factors, with positive associations for median household income and racial diversity, alongside negative associations with physical boundaries and symbolic US flag. The spatial model revealed additional associations with greenery and building. These patterns are consistent with selection processes whereby agreeable individuals, who value social harmony, may prefer neighborhoods characterized by economic resources and social diversity rather than areas marked by physical barriers. The role of racial diversity aligns with research suggesting that socially diverse environments attract and reinforce prosocial tendencies \citep{laceulle_transactions_2018}.

Openness had the highest overall model fit (R$^{2}$ = 0.467) but was driven predominantly by demographic age composition rather than built environment features. All four age group shares were highly significant predictors, while no grouped environmental category reached conventional significance. This suggests that the geographic variation in Openness may be better explained by who lives in a neighborhood—particularly the age structure of its residents—than by the physical characteristics of the environment itself. These age shares may further act as proxies for unobserved neighborhood features—such as proximity to universities, concentrations of high-skill or creative-class employment, and entertainment districts—that both attract younger residents and correlate independently with Openness, so the age effect should not be read as a direct effect of age per se. Age-based selection processes, such as younger cohorts with higher Openness concentrating in particular neighborhoods, may be the dominant mechanism for this trait.

Extraversion and Neuroticism showed the fewest built environment associations. Extraversion was primarily linked to socioeconomic characteristics (income positively, racial diversity negatively), while Neuroticism's main predictor was population density. The positive association between density and Neuroticism could reflect either selection (more neurotic individuals being constrained to denser, potentially more affordable areas) or environmental influence (denser environments generating stress that reinforces negative emotionality).

These selection and environmental influence processes likely operate simultaneously and reinforce each other over time, creating feedback loops where personality distributions and built environment features mutually shape each other. Personality traits also influence civic participation in urban planning \citep{li_personality_2025}, suggesting that neighborhoods with particular personality concentrations may develop distinct environmental characteristics through collective decision-making. However, disentangling these mechanisms requires longitudinal data tracking both personality changes and residential mobility, as well as comparisons between long-term residents and newcomers. Future research should prioritize such designs to establish the relative contributions of each mechanism.

The explanatory power of our models merits consideration. Our Openness model explained 46.7\% of variance in the OLS specification (47.9\% in the SLM), which is notable but not unprecedented. \citet{jokela_geographically_2015} found that neighborhood-level variables explained up to 78\% of variance in Openness across 216 London postal districts using backward stepwise regression with census-based predictors. Several differences help explain why our R$^{2}$ values are lower: our predictor set draws on SVI-derived visual features rather than the census-based socioeconomic and demographic variables used in prior work, our theory-driven grouping with VIF-based selection avoids the overfitting that backward stepwise procedures are known to inflate, and our ZCTA unit of analysis is coarser than a London postal district. The coefficients and significance patterns reported here should therefore be read as lower bounds on what a richer feature set and finer spatial unit might recover. Even so, the substantial explanatory power across traits—particularly for Openness and Agreeableness (R$^{2}$ = 0.353)—supports the conclusion that personality trait distributions are meaningfully structured by neighborhood-level characteristics, reinforcing earlier findings at the postal-district scale. The modest magnitude of the individual bivariate correlations and coefficients we report is also consistent with effect sizes commonly observed in psychological research, where small effects accumulate into meaningful differences at the population level \citep{gotz_small_2022, gotz_effect_2024}.

\subsection{Implications for Geography Studies}

This study advances geography studies in several important ways. First, it integrates fine-scale analysis of personality traits into the field of geography along with advanced geocomputing methods for the first time. By combining personality traits with geographic analysis, we demonstrate how individual psychological differences can be spatially distributed and associated with environmental features, opening new avenues for understanding human-environment interactions.

Second, our approach advocates for more human-centered perspectives in geography and GeoAI. Rather than focusing solely on aggregate demographic or economic patterns, we examine how psychological characteristics—fundamental aspects of human individuality—relate to place characteristics. This perspective recognizes that places are not just containers for human activity but are experienced and potentially shaped by the personality traits of people.

Third, methodologically, this research demonstrates the potential of combining GeoAI approaches with psychological data to generate new insights about human geography. Our use of computer vision to analyze street view imagery at scale with fine-resolution personality data represents an innovative approach that could be applied to other psychological constructs and geographic contexts.

Finally, our findings suggest that geographic patterns of personality may be more pronounced and systematic than previously recognized, particularly at neighborhood scales. This has important implications for understanding social and economic processes, as personality traits influence numerous life outcomes, including health, economic behavior, and social relationships. Geographic clustering of personality traits may therefore contribute to regional differences in various social and economic phenomena.

\subsection{Practical Implications}

While our findings reveal correlational associations rather than causal relationships, they offer useful information for urban planners and policymakers. Understanding the personality trait distributions across neighborhoods can help anticipate potential social outcomes and characteristics, given the well-established links between personality traits and various societal phenomena such as health behaviors, civic engagement, and economic activity. For instance, planners working in areas with particular personality profiles might better understand local community needs and preferences, informing decisions about public space design, community programming, or infrastructure priorities. Similarly, this information could help explain existing patterns in neighborhood characteristics such as social cohesion, volunteering rates, or local economic vitality.

Looking forward, once causal relationships between built environment features and personality traits are established through longitudinal and intervention studies, these insights could enable more targeted urban design strategies. Such evidence would allow planners to implement concrete interventions aimed at fostering specific psychological and social outcomes. However, until causality is firmly established, the current correlational evidence serves primarily to deepen our understanding of existing neighborhood dynamics and to generate hypotheses for future research that could ultimately inform evidence-based urban planning practices.

\subsection{Generalizability and Contextual Considerations}

Our study focuses on four cities in Texas, a decision that involves an important trade-off between internal validity and external generalizability. By selecting cities that share similar governance structures, climate conditions, cultural contexts, and urban development patterns, we reduce confounding factors that might obscure personality-environment relationships. This regional similarity strengthens our ability to isolate genuine associations between built environment features and personality traits. However, this same homogeneity limits our ability to generalize findings to cities with different characteristics, such as compact transit-oriented cities, different cultural contexts, or varied climatic conditions. Environmental features may carry different meanings and relate differently to personality traits in contexts that differ from the sprawling, car-centric urban form characteristic of Texas cities.


An open question remains about the optimal scale at which to embrace this trade-off. While studying highly similar contexts strengthens internal validity, we do not yet know whether personality-environment associations operate similarly at regional, national, or global scales, or whether they are fundamentally context-dependent. Future research should explore this question by conducting comparative studies across cities with varying degrees of similarity. Such investigations would help identify which personality-environment associations represent universal patterns versus context-specific phenomena, ultimately building a more complete understanding of how human psychological traits relate to built environments across diverse urban settings worldwide.

\subsection{Ethical Considerations}

Research linking personality traits to geographic locations raises important ethical considerations. While our analyses use aggregated ZCTA-level data that protects individual privacy, the identification of personality-environment associations could potentially be misused for discriminatory purposes, such as profiling neighborhoods or making assumptions about residents based on their location. We emphasize that our findings describe statistical associations at the aggregate level and should not be used to infer the personality characteristics of individual residents---a risk known as the ecological fallacy \citep{robinson_ecological_1950}.

The personality data used in this study were voluntarily provided by participants through an online assessment platform, and geographic information was limited to self-reported ZIP codes. Our aggregation to the ZCTA level, combined with the minimum threshold of 20 participants per ZCTA, provides an additional layer of privacy protection. Nevertheless, as research in geographic psychology advances and spatial resolution improves, the research community should develop clear ethical guidelines for the responsible use of spatially referenced psychological data, ensuring that such information serves to improve urban environments rather than reinforce stereotypes or enable discrimination. We note that regional variation in personality is already well documented in the literature we build on \citep{rentfrow_geographical_2013, ebert_are_2022}, so our contribution is methodological rather than the introduction of a new category of sensitive information. The same knowledge also supports beneficial uses, such as identifying neighborhoods more susceptible to psychological stressors or tailoring public-health and planning interventions to local needs. The appropriate response is therefore to pair reporting with clear guidance on interpretation---particularly the ecological fallacy and the modest magnitude of the associations---rather than to withhold descriptive findings.

\subsection{Limitations and Future Directions}
Several limitations of our approach are acknowledged, and potential future research directions are highlighted here.
First, our cross-sectional design precludes examination of the temporal dimension of personality-environment relationships. While individual personality traits remain relatively stable across adulthood, the aggregate personality of a neighborhood can still change over time as residents move in and out and as demographic composition shifts, and built environments themselves change through development and revitalization. This limitation opens several research opportunities: Longitudinal studies could track both personality distributions and built environment changes over time to reveal how these relationships evolve and establish bidirectional causality (i.e., personality → built environment and built environment → personality). Large-scale environmental interventions designed as experiments are rarely feasible in practice, so natural experiments that exploit unanticipated environmental change—such as the decline of atmospheric lead following the phase-out of leaded fuel \citep{schwaba_impact_2021}—offer a more realistic path for causal inference. Additionally, comparing the personalities of long-term residents versus newcomers could help clarify whether observed associations stem from selective migration or environmental influence. A more data-driven approach could also employ causal discovery and inference methods with many geographic and demographic variables to identify the most relevant predictors of personality traits \citep{feng_causal_2025}. Once such causal links are confirmed, more targeted interventions could be designed to modify built environments in ways that promote positive psychological outcomes.

Second, while our study examines associations across four Texan cities, our analysis was limited to ZCTA-level aggregations and visual street view data. Restricting the sample to four large metropolitan areas also narrows the range of built environment variation relative to the full spectrum of rural, small-town, and non-Texan urban contexts, which likely attenuates the magnitude of the observed associations. These constraints may limit generalizability to other geographical and cultural contexts, potentially obscure finer-grained variations within neighborhoods, and may not capture all sensory experiences of place, such as sounds, smells, or tactile qualities that may also influence personality-environment relationships. These limitations suggest several promising directions: cross-cultural studies could examine whether the patterns we identified are universal or culturally specific. More fine-grained spatial analyses at the block or street level might reveal more nuanced relationships. Integration of other sensory data beyond visual features could provide a more holistic understanding of environment-personality relationships. Finally, research exploring the psychological mechanisms underlying these associations—such as stress responses, attentional patterns, or social interaction dynamics—would significantly advance our understanding of how physical spaces and personality traits interact, building on findings by studies such as \citet{coutrot_entropy_2022} who find effects of street network entropy on human cognition.

Lastly, we encourage future studies to apply this methodology to a broader range of geographic areas—including rural and suburban contexts in addition to cities—across different states, countries, and societies with diverse social, economic, and historical contexts. Scaling up such analyses could yield valuable insights into how built environments interact with personality traits across varied geographies and populations.

\section{Conclusions}
\label{sec:conclusions}

This study explored associations between Big Five personality traits and built environment characteristics derived from street view imagery across four Texas cities. Our findings reveal significant spatial clustering of personality traits at the ZCTA level, with meaningful correlations between grouped environmental features and trait distributions. Openness showed the strongest overall model fit in our OLS models (R$^{2}$ = 0.467), driven primarily by demographic age composition, while Conscientiousness exhibited the most extensive associations with built environment features, with five of eight grouped categories and symbolic US flag reaching at least marginal significance. Agreeableness (R$^{2}$ = 0.353) was shaped by socioeconomic factors and select built environment features, including physical boundaries and symbolic US flag. These relationships are correlational rather than causal, likely reflecting complex underlying processes and potentially driven by third variables such as socioeconomic factors or historical development patterns.

These findings contribute to geography and GIS studies by revealing relationships between psychological traits and physical spaces at the neighborhood level. The theoretical implications for geography are substantial, as understanding how environmental features relate to different personality types can inform geographic theories about human-environment interactions and spatial patterns of human behavior. Geographic analysis of personality traits represents a new frontier that bridges psychological and spatial perspectives, offering insights into how individual differences manifest spatially and relate to place characteristics. By integrating personality with spatial analysis and GeoAI methods, researchers can develop a richer understanding of how personality traits and the built environment shape each other, ultimately contributing to more human-centered approaches in geography studies and spatial analysis.

\section*{Acknowledgments}
This research was funded by the Singapore International Graduate Award (SINGA) scholarship provided by the Agency for Science, Technology, and Research (A*STAR).
This research is part of the project Large-scale 3D Geospatial Data for Urban Analytics, which is supported by the National University of Singapore under the Start Up Grant R-295-000-171-133.
Y.K. acknowledges the funding support provided by the Population Research Center Grant P2CHD042849, NICHD.
The content is solely the responsibility of the authors and does not necessarily represent the official views of the National Institutes of Health.

\section*{Declaration of generative AI and AI-assisted technologies in the writing process}
During the preparation of this work, the author(s) used Claude Sonnet 3.7 to format tables and proofread. After using this tool/service, the author(s) reviewed and edited the content as needed and take(s) full responsibility for the content of the published article.

\newpage
\appendix

\section{Segmentation Variables}

\begin{table}[htbp]
\centering
\small
\caption{Grouped built environment categories and their constituent variables from street view imagery analysis.}
\label{tab:segmentation_variables}
\begin{tabularx}{\textwidth}{llX}
\toprule
\textbf{Category} & \textbf{Type} & \textbf{Constituent Variables} \\
\midrule
\multicolumn{3}{l}{\textit{Grouped categories (composite scores)}} \\
Greenery & Ratio & Vegetation \\
Open Space & Ratio & Sky, Terrain, Sand \\
Building & Ratio & Building \\
Road & Ratio & Road, Service Lane \\
Active Mobility Infrastructure & Ratio & Sidewalk, Bike Lane, Pedestrian Area \\
Active Mobility Presence & Count & Person, Bicycle, Bicyclist \\
Vehicle Presence & Count & Car, Truck, Bus, Motorcycle, Trailer, Other Vehicle, Caravan \\
Physical Boundaries & Ratio & Fence, Wall, Barrier \\
\midrule
\multicolumn{3}{l}{\textit{Individual features (not grouped)}} \\
Symbolic US Flag & Count & American Flag \\
CCTV Surveillance & Count & CCTV Camera \\
Visual Complexity & --- & Shannon entropy of all segmentation pixel ratios \\
\bottomrule
\end{tabularx}
\begin{flushleft}
\footnotesize
Ratio variables are derived from semantic segmentation (pixel area proportions); count variables from object detection. For each grouped category, constituent variables were z-scored and summed to create composite scores, then re-standardized. Individual features were z-scored directly.
\end{flushleft}
\end{table}

\section{Temporal Distribution of Data Sources}

\begin{figure}[htbp]
\centering
\includegraphics[width=\textwidth]{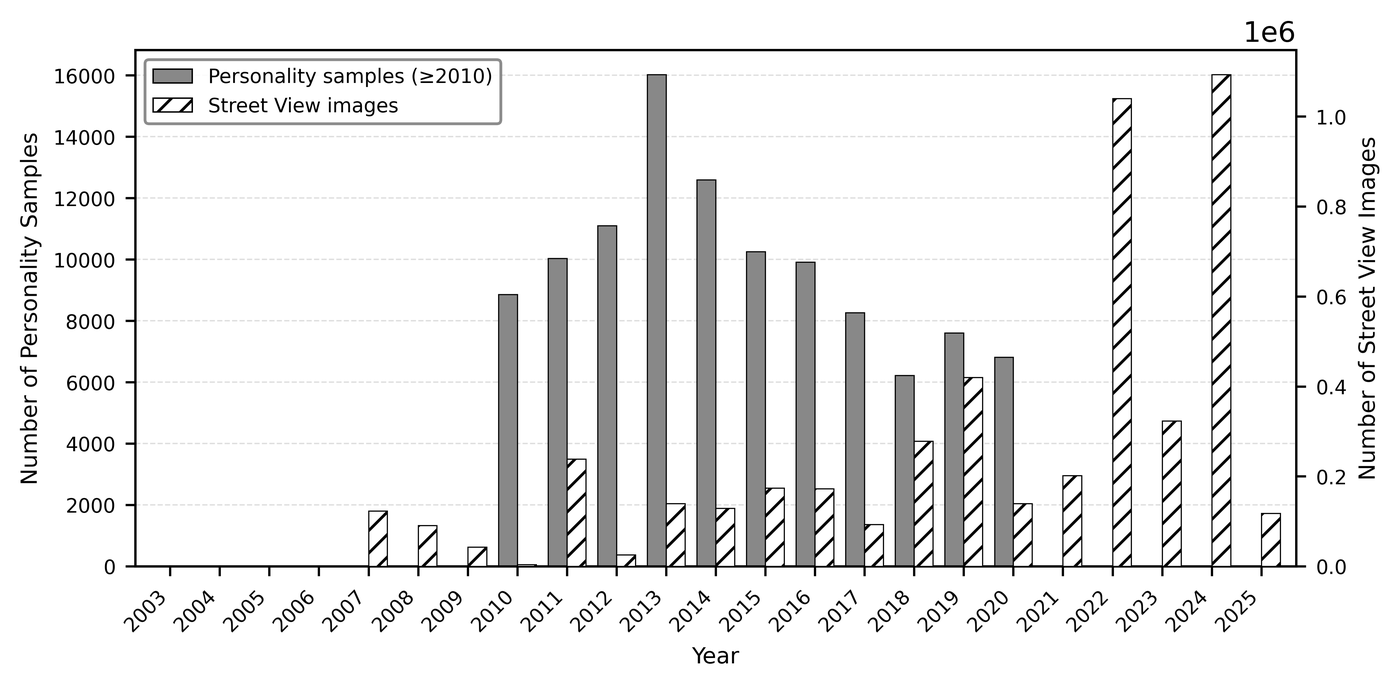}
\caption{Temporal distribution of personality survey samples and Google Street View images across the study period. Gray bars indicate the number of personality survey responses per year (filtered to 2010 onward); hatched bars indicate the number of street view images collected per year.}
\label{fig:temporal_histogram}
\end{figure}

\section{ZIP Code Model Inclusion}

\begin{figure}[htbp]
\centering
\includegraphics[width=\textwidth]{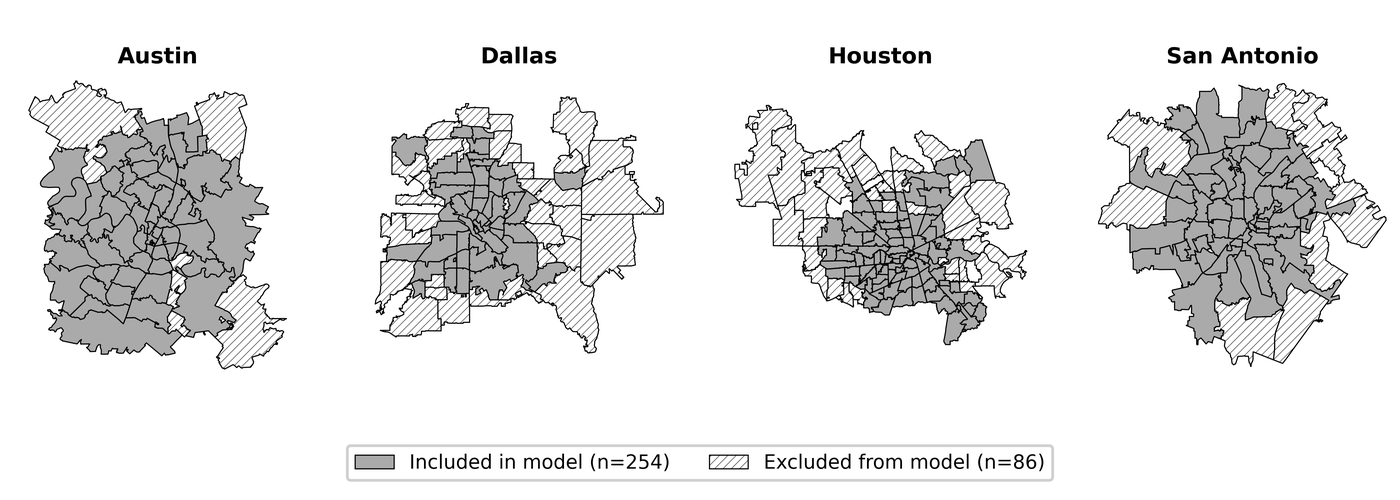}
\caption{ZCTA model inclusion across the four study cities. Gray areas indicate ZCTAs included in the regression analysis; hatched areas indicate ZCTAs excluded due to insufficient personality survey responses (fewer than 20 participants) or inadequate street view imagery coverage.}
\label{fig:zipcode_inclusion}
\end{figure}

\newpage
\bibliographystyle{cas-model2-names}





\end{document}